\documentclass[usenatbib]{mn2e}
\usepackage{graphicx}
\usepackage{amsmath}

\def\apj{\rm ApJ}
\def\apjl{\rm ApJL}
\def\apjs{\rm ApJS}

\def\aaps{\rm A\&AS}
\def\mnras{\rm MNRAS}
\def\nat{\rm Nature}

\def\aap{\rm AAP}
\def\araa{\rm ARA\&A}

\def\procspie{\rm Proc. SPIE}

\def\gax{\mathrel{\raise.3ex\hbox{$>$}\mkern-14mu\lower0.6ex\hbox{$\sim$}}}
\def\lax{\mathrel{\raise.3ex\hbox{$<$}\mkern-14mu\lower0.6ex\hbox{$\sim$}}}
\def\gtorder{\mathrel{\raise.3ex\hbox{$>$}\mkern-14mu
             \lower0.6ex\hbox{$\sim$}}}
\def\ltorder{\mathrel{\raise.3ex\hbox{$<$}\mkern-14mu
             \lower0.6ex\hbox{$\sim$}}}

\newcommand{\thisevent}{SN~2012fh}
\begin{document}

\title [Progenitor of Supernova 2012fh]{On the Progenitor of the Type Ibc Supernova 2012fh}

\author[Samson~A.~Johnson]{ 
    Samson~A.~Johnson$^{1}$,
    C.~S.~Kochanek$^{1,2}$,
    S.~M.~Adams$^3$
    \\
  $^{1}$ Department of Astronomy, The Ohio State University, 140 West 18th Avenue, Columbus OH 43210 \\
  $^{2}$ Center for Cosmology and AstroParticle Physics, The Ohio State University, 191 W. Woodruff Avenue, Columbus OH 43210 \\
  $^{3}$ Cahill Center for Astrophysics, California Institute of Technology, Pasadena, CA 91125\\
   }

\maketitle

\begin{abstract}
Little is observationally known about the progenitors of Type Ibc supernovae (SNe) or the typical activity of SNe progenitors in their final years. 
Here, we analyze deep Large Binocular Telescope imaging data spanning the 4 years before and after the Type Ibc \thisevent~using difference imaging. 
We place 1$\sigma$ upper limits on the detection of the progenitor star at $M_U>-3.8$, $M_B>-3.1$, $M_V>-3.8$, and $M_R>-4.0$ mag.
These limits are the tightest placed on a Type Ibc SNe to date and they largely rule out single star evolutionary models in favor of a binary channel as the origin of this SN. 
We also constrain the activity of the progenitor to be small on an absolute scale, with the RMS \textit{UBVR} optical variability $\lax2500\text{L}_\odot$ and long-term dimming or brightening trends $\lax1000\text{L}_\odot/\text{year}$ in all four bands.
\end{abstract}


\section{Introduction}
\label{sec:introduction}

The observed variety of core collapse supernovae (ccSNe) implies differences in their progenitor systems. 
In particular, ccSNe are placed broadly into two categories, Type I and Type II, based on the absence or presence of Hydrogen lines in their explosion spectra. 
Type~II ccSNe are classified further by the structure of their light curves. 
Type~I ccSNe are sub-divided into Type~Ib and Ic based on the presence or absence of Helium emission lines (\citealt{fili1997}). 

The progenitors of Type II ccSNe have been identified as red supergiants through direct imaging \citep[see the review by][]{smartt2009}.
However, there is no definitive detection of the progenitor to a Type~Ib or Type Ic (hereafter Type Ibc) ccSN. 
The progenitors of Type Ibc ccSNe are believed to be stripped Wolf-Rayet (WR) stars.
They can be stars that began their lives with large initial masses ($M_\textrm{ini}\gax25\textrm{M}_\odot$) which lose a significant amount of mass through strong stellar winds, or they can be stars stripped by mass loss in an interacting binary \citep[e.g.,][]{eldridge2008}.
A possibly related puzzle is the near absence of detected ccSNe progenitors with masses $\gax17 \textrm{M}_\odot$ \citep{kochanek2008,smartt2009}.
One possibility is that these more massive stars all evolve to WR stars that explode as Type Ibc ccSNe \citep[e.g.,][]{groh2013a}.
WR stars are extremely luminous but optically faint because almost all their energy is radiated in the ultraviolet, and are thus difficult to detect. 
Alternatively, this could be evidence for failed SNe, where a black hole is formed without an explosion \citep{kochanek2008}.
Detections or strong constraints on the progenitors of Type Ibc ccSNe are crucial for understanding these puzzles.

\begin{figure*}
\includegraphics[width=\linewidth]{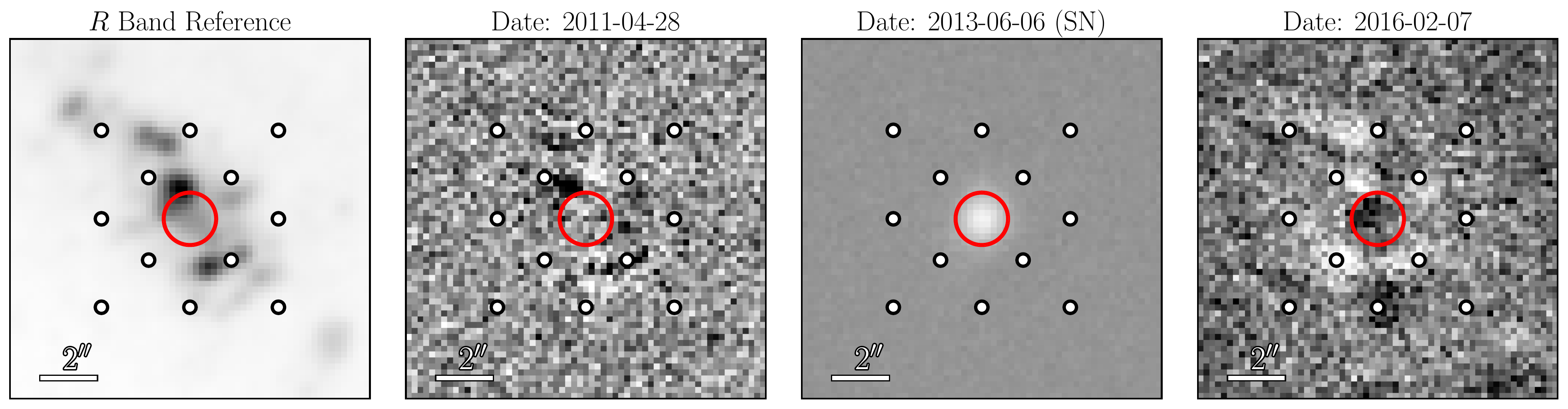}
\caption{
\textit{R}-band images centered on the location of \thisevent. From left to right, the panels are the reference image, a subtracted image prior to the SN, a subtracted image with the SN, and a subtracted image at much later times. All images are on a linear color scale, with the scale for the subtracted images being symmetric about zero. 
The subtracted images are scaled such that deficits in flux relative to the reference image are darker and excesses are whiter. 
The red circles are 1$^{\prime\prime}$ in radius. 
The twelve blue circles indicate the positions we use for our comparison sample.
The progenitor should appear as a dark point source at the center of the red circle in the right panel. 
}
\label{fig:4panelR}
\end{figure*}

\citet{eldridge2013} reviewed 12 Type~Ibc ccSNe that had archival \textit{Hubble Space Telescope} (\textit{HST}) images of their host galaxies taken prior to the explosion.
They derived magnitude limits for each, of which the strongest was for the progenitor of SN 2002ap \citep{crockett2007} at $M_B\geq-4.2$ and $M_R\geq-5.1$ mag.
\citet{eldridge2013} argue that the limits imply that the dominant channel for producing Type Ibc ccSNe must be binary evolution.
A candidate stripped progenitor to the Type Ib SN iPTF13bvn was discovered by \citet{cao2013}, for which \citet{groh2013a} and \citet{eldridge2015} suggest a single and binary progenitor system, respectively.
Observations after the SN fades can confirm the identity of the progenitor and clarify the evolutionary scenario.

The behavior of ccSNe progenitors in their final years is also a current topic of debate. 
Some appear to be quiescent in their golden years \citep{szczygiel2012,fraser2014,kochanek2017}, while others exhibit eruptive events \citep[e.g.,][]{pastorello2007,fraser2013,mauerhan2013,ofek2016} or show evidence for significant pre-SN mass loss through interactions with a dense circumstellar medium \citep[e.g.,][]{galyam2012,margutti2017,yaron2017}.
Understanding this problem requires measuring the variability of SNe progenitors with well determined or constrained properties. 
This requires measurements with greater sensitivity than typical SNe surveys. 

We are monitoring 27 nearby galaxies to search for failed SNe using the Large Binocular Telescope \citep[LBT,][]{kochanek2008,gerke2015,adams2017}.
These data also allow us to study the luminosities, temperatures, and variability of progenitors to successful SNe in these galaxies \citep{szczygiel2012, kochanek2017}. 
Here we examine the progenitor of the Type~Ibc \thisevent. 
\thisevent~was discovered by \citet{nakano2012} on 2012-10-18 in the galaxy NGC~3344 at RA=10:43:34.05, Dec=24:53:29.00 and was classified as Type Ic by \citet{tomasella2012} and \citet{takaki2012}.
They estimated the initial detection was $\sim130$ days after the explosion. 
Because the spectra were obtained long after peak, we will be conservative and refer to the event as Type Ibc. 
The SN was not observed at its peak due to the Sun.
As noted in \citet{gerke2015}, \thisevent~was also present in the LBT survey data.
 
In this paper, we present deep LBT photometry of the progenitor location leading up to the explosion and then as the SN fades. 
In Section~\ref{sec:observations}, we detail the observations of the host galaxy and our procedure for extracting data. 
We set limits on the luminosity and variability of the progenitor in Section~\ref{sec:progenitor}.
Finally, we conclude with a discussion of our findings in Section~\ref{sec:discussion}. 
For this analysis, we adopt a distance to the host galaxy NGC 3344 of 6.9 Mpc \citep{verdes2000}, a Galactic extinction of $E(B-V)=0.0281$ mag for an $R_V=3.1$ reddening law \citep{schlafly2011}.
The \textit{Swift} UV fluxes found by \citet{margutti2012} imply little extinction local to the SN. 

\section{Observations}
\label{sec:observations}

The images of the host galaxy were obtained using the Large Binocular Camera (\citealt{giallongo2008}) on the LBT (\citealt{hill2006}).
Our reduction and subtraction procedures are identical to those of \citet{gerke2015} and \citet{adams2017} except for the images used to construct the reference frame.
We use the image subtraction software \texttt{ISIS} (\citealt{alard1998}, \citealt{alard2000}) for the analysis and PSF photometry, aligning all the data to a common astrometric solution in all four filters. 

\begin{figure}
\includegraphics[width=\linewidth]{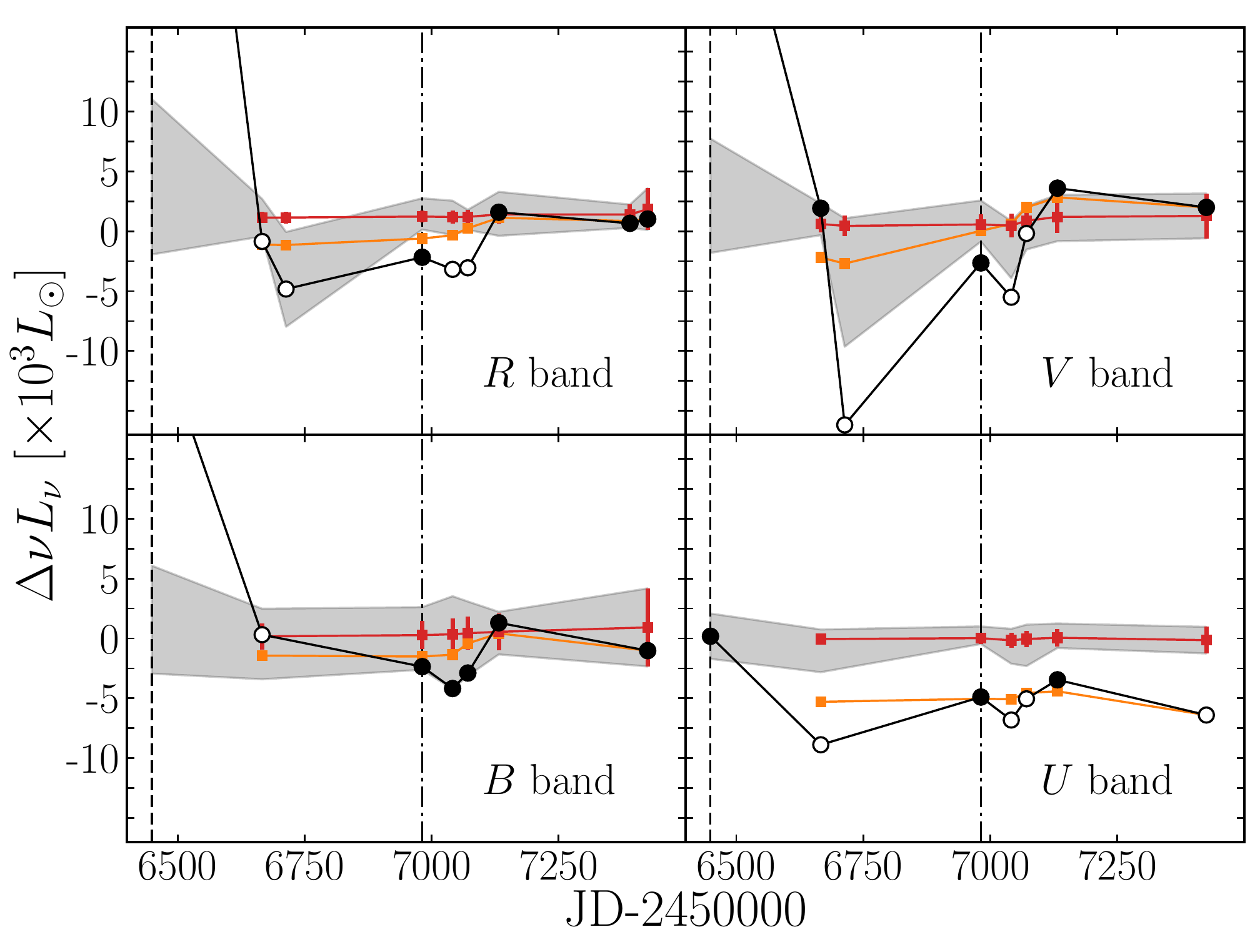}
\caption{
Differential photometry for the period following \thisevent. 
The luminosities $\nu L_\nu$ at the position of the SN are shown in black, with open symbols indicating lower quality epochs with seeing above $1\farcs5$ or a flux scaling factor $<0.8$.
The orange points are a running average of these points moving backwards in time from the most recent epoch.
The red values are the moving average of the mean of the comparison sample and the gray regions depict the 1$\sigma$ dispersion about this mean for each epoch.
The vertical dashed line marks the epoch (2016-06-06) containing \thisevent~in the LBT data. 
We adopt the luminosities at the epoch indicated by the dot-dash line as our 1$\sigma$ limit on the progenitor.
}
\label{fig:4post}
\end{figure}

\begin{figure*}
\includegraphics[width=\linewidth]{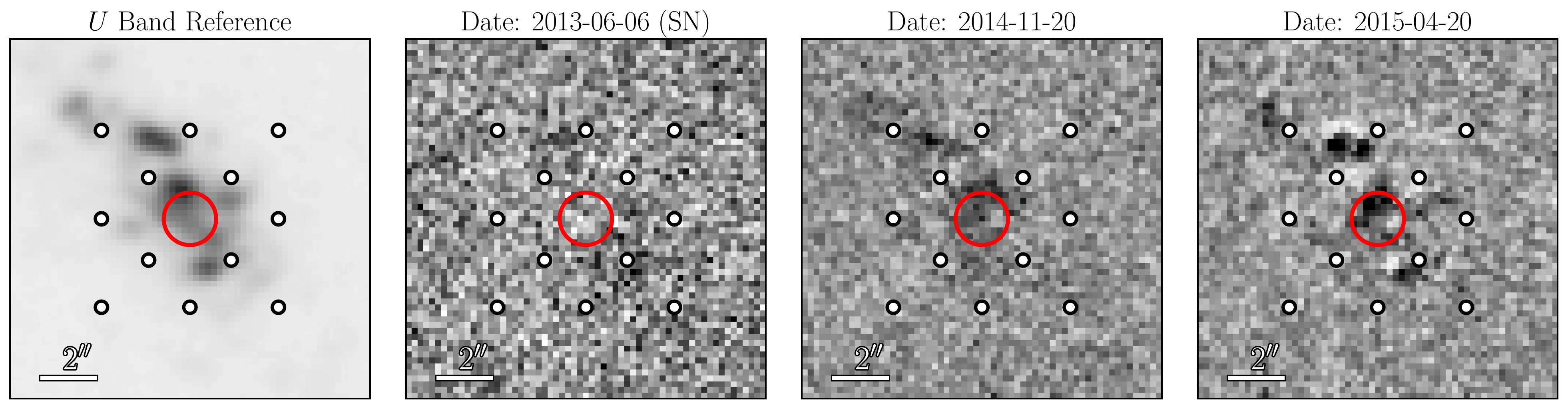}
\caption{
\textit{U}-band images centered on the location of \thisevent. 
The left panel shows the reference image, followed by the subtracted image at the epoch with the \textit{R} band detection of the SN, and two later epochs taken in good conditions. 
The progenitor would appear as a dark point source at the center of the $1^{\prime\prime}$ red circle if it were detected.
Instead, the subtraction residuals from the nearby sources in the two right panels drive the \textit{U} band ``detection'' seen in Figure \ref{fig:4post}.
}
\label{fig:4panelU}
\end{figure*}

We assemble our reference frames using only images obtained prior to the SN.
This way, the post-explosion difference images show the deficit of light from the absence of the progenitor once the SN fades.  
The subtracted pre-explosion images should reflect any variability about the average luminosity of the progenitor in the $\sim$4 years prior to the explosion. 
The reference frames are assembled from the best pre-SN images: those with $\lax1\farcs3$ seeing, low background, and no evidence of clouds/cirrus.
The \textit{UBVR} reference images are comprised of  5, 6, 6, and 14 images, respectively. 
The zero points of our reference frames are determined using SDSS photometry of the field \citep{ahn2012}, and we convert from \textit{ugriz} to \textit{UBVR} photometry using the procedure described by \citet{jordi2006}. 
While we show results for all the data included in our analysis, we also flag ``low quality'' data defined by seeing $>1\farcs5$ or the \texttt{ISIS} flux scaling factor being $<0.8$.
A low flux scaling factor indicates that the image either was taken through cirrus or at a significantly higher than average airmass.  

The LBT data taken on 2013-06-06 contains the SN, allowing us to accurately determine the position of the progenitor. 
The position was fixed to the centroid of the SN on the \textit{R} band image.
All the data has been interpolated to a common astrometric reference, making this position the same for all other filters. 
As seen in the leftmost panel of Figure \ref{fig:4panelR}, \thisevent~was located near a cluster of bright stars.
This results in larger subtraction residuals than would be found given a smoother background.
We place a grid of 12 sample points around the position of the SN for later comparison.
The grid spacing of the outer points is 15 pixels ($\sim$3\farcs5 given the $0\farcs2255$ pixel$^{-1}$ scale), and the inner grid spacing is 7 pixels.
The positions are displayed as circles in Figure \ref{fig:4panelR}.
By comparing the photometry of the progenitor to that of our sample points, we can better understand any systematic errors in the light curve.
We extract light curves at the position of the SN and the comparison sample using the standard PSF-weighted estimates produced by \texttt{ISIS}.  

\begin{table}
\caption{Detection limits.}
\begin{tabular}{c c c c r}
\multicolumn{1}{c}{Band}
&\multicolumn{1}{c}{Apparent}
&\multicolumn{1}{c}{Absolute}
&\multicolumn{1}{c}{Luminosity}
&\multicolumn{1}{c}{Aperture}\\
\multicolumn{1}{c}{}
&\multicolumn{1}{c}{[mag]}
&\multicolumn{1}{c}{[mag]}
&\multicolumn{1}{c}{$[\nu L_\nu/\text{L}_\odot]$}
&\multicolumn{1}{c}{$[\nu L_\nu/\text{L}_\odot]$}\\
\hline
$R$&$>25.2$& $>-4.0$ & $ <2200$ & $700$ \\
$V$&$>25.4$& $>-3.8$ & $ <2600$ & $-800$ \\
$B$&$>26.1$& $>-3.1$ & $ <2400$ & $-200$ \\
$U$&$>25.4$& $>-3.8$ & $ <4900$ & $4600$ \\

\end{tabular}
\label{tbl:mag_limits}
\end{table}

\begin{figure}
\includegraphics[width=\linewidth]{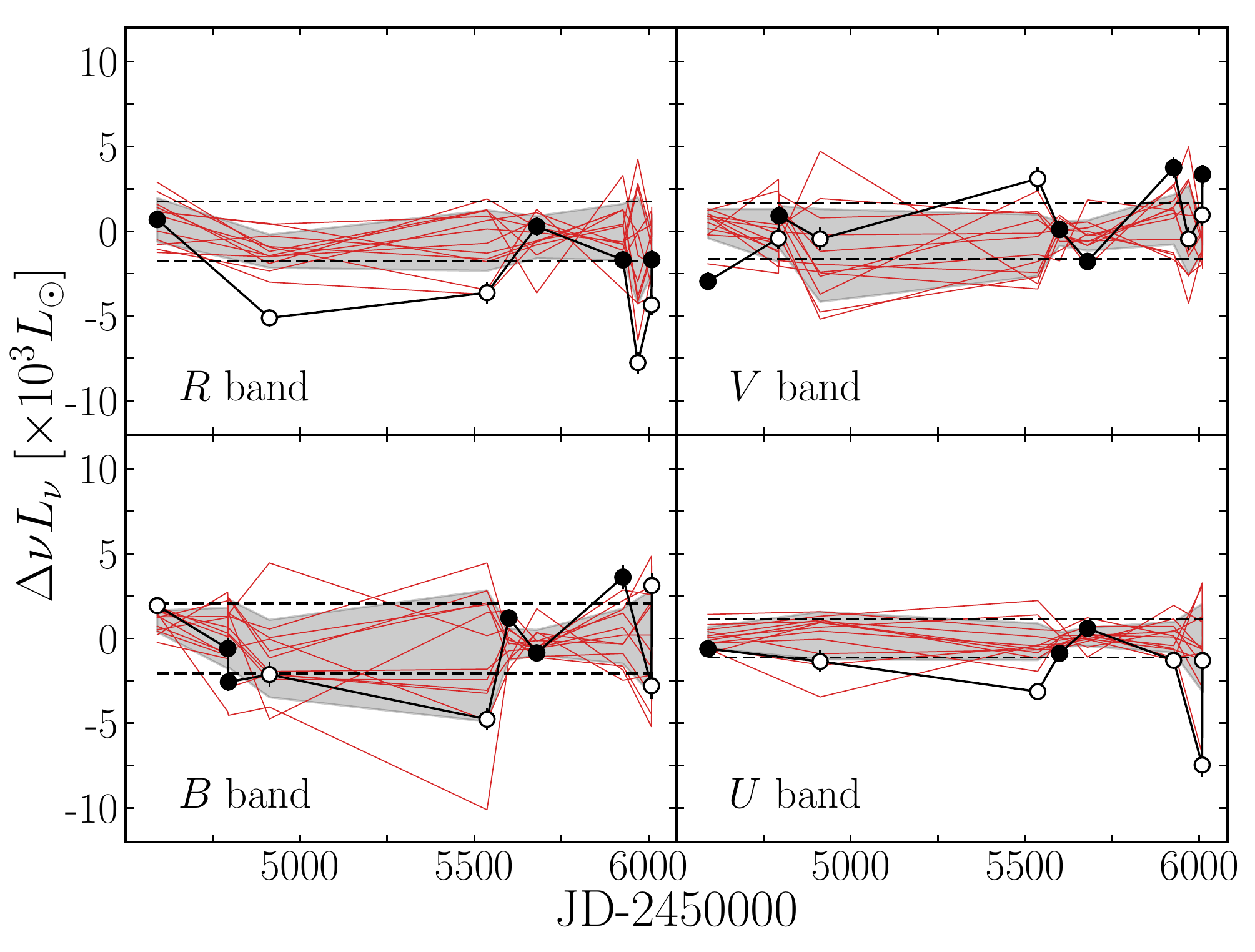}
\caption{
Differential photometry of \thisevent~progenitor (black points). 
The open points correspond to poorer quality data.
The red lines are the light curves for the comparison grid shown in Figure \ref{fig:4panelR}. 
The gray region depicts the 1$\sigma$ boundary about the mean of these light curves for each epoch.
The black dashed lines indicate the mean root-mean-square values of the comparison light curves across all pre-SN epochs.
The observed scatter in the luminosity of the progenitor is consistent with the comparison sample, indicating no detection of pre-SN variability for \thisevent. 
}
\label{fig:4pre}
\end{figure}

\section{Limits on the Progenitor}\label{sec:progenitor}
To place limits on the luminosity of the progenitor, we use the difference images following \thisevent. 
Since we built the reference image using only images prior to the SN, post-SN difference images should show a deficit with a point source flux equal to the luminosity of the progenitor once the SN has faded. 
We see no evidence of emission from the SN in any epoch after the 2013-06-06 epoch shown in Figure \ref{fig:4panelR}.

We perform a weighted moving average from the last post-explosion subtracted image and moving backwards through the epochs. 
This could typically be used to determine the luminosity of a progenitor as a SN fades, but no signal is apparent in Figure~\ref{fig:4post}.
We also show the weighted moving average of the mean luminosity of the comparison sample. 
The gray shaded region depicts the 1$\sigma$ scatter about this mean. 
There appears to be a deficit in the \textit{U} band light curve, which could be interpreted as flux from the progenitor.
However, as shown in Figure \ref{fig:4panelU}, the \textit{U} band flux appears to be due to subtraction residuals from nearby sources and there is no point-like source centered on the position of the SN.
Hence, we interpret this only as a conservative upper limit.
We adopt the fluxes measured for the ``good quality'' observations on 2014-11-20 as $1\sigma$ limits for the detection of the progenitor, indicated in Figure~\ref{fig:4post} with a dot-dash line. 
This choice is broadly consistent with any other good epoch, and is slightly more conservative than the moving averages. 
The fluxes, their conversion into absolute magnitudes, and the band luminosities $(\nu L_\nu)$ are provided in  Table~\ref{tbl:mag_limits}.

We also performed aperture photometry on two types of stacks of post-SN difference images.
For the first stack, we compute the average of all post-SN images.
The second stack is the average of all ``good'' quality post-SN images. 
We first perform aperture photometry on the location of the progenitor and three bright point sources with a signal aperture radius of 3 pixels and a sky annulus with an inner and outer radii of 16 and 20, respectively.
We then use a signal aperture of 9 pixels around the bright sources to compute an aperture correction.
The resulting estimate of the flux at the SN location from both averages of the images are similar.
We average these two results and report them in Table \ref{tbl:mag_limits} as ``Aperture''.
To remain conservative, we maintain the limits from the best epoch.


\begin{figure}
\includegraphics[width=\columnwidth]{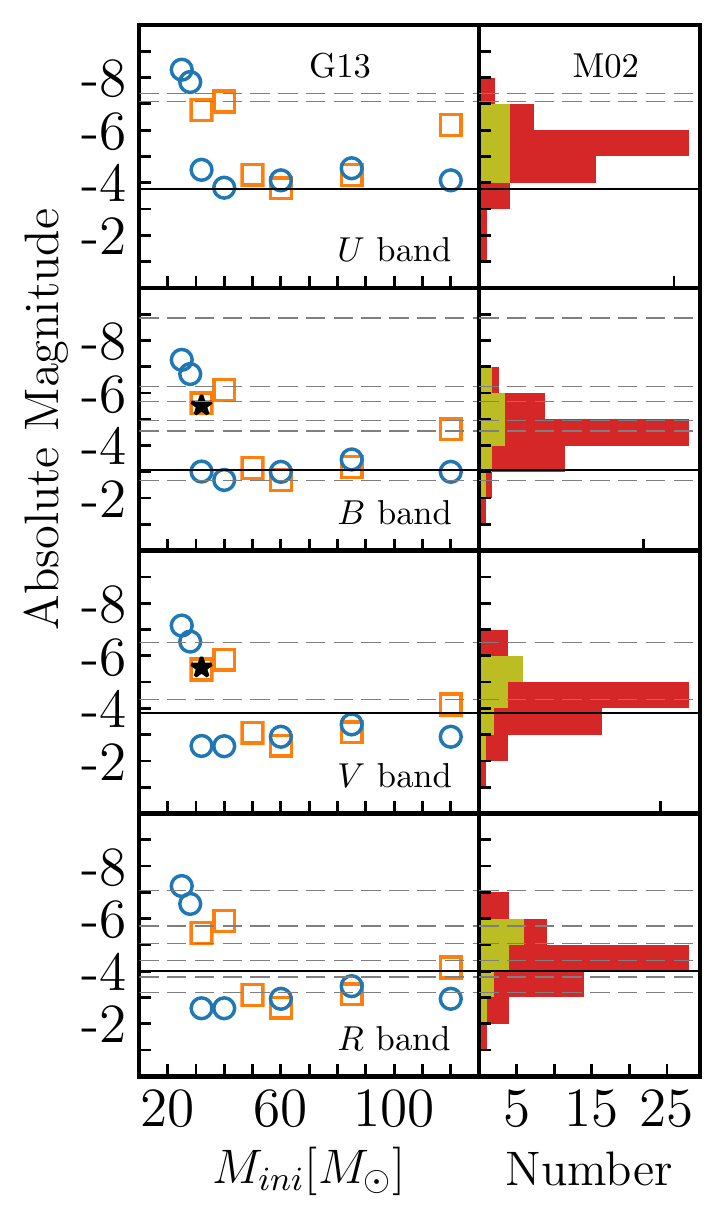}
\caption{
Comparison of our $1\sigma$ progenitor detection limits to model and observed absolute magnitudes of single Wolf-Rayet stars.
The left panels show the model magnitudes from \citet[][G13]{groh2013b} as orange squares (blue circles) for the non-rotating (rotating) models.
The shaded gray regions mark the estimated range of the iPTF13bvn progenitor magnitudes by \citet{folatelli2016}. 
The right panels show the comparison sample of LMC single WR stars from \citet{massey2002} used by \citet{eldridge2013}. 
WN type stars are represented by red bars, and WO and WC are blue.
The black horizontal lines indicate our $1\sigma$ limit for each band, while the gray dashed lines are the limits on Type Ibc progenitors compiled by \citet{eldridge2013} rescaled to be $1\sigma$ limits (instead of $5\sigma$).
This limit on the \textit{U} band limit is the strongest placed on a Type Ibc SN progenitor, and largely excludes many of the model and observed WR stars.
}
\label{fig:limits}
\end{figure}

In Figure \ref{fig:limits} we compare our limits to model single star progenitors from \citet{groh2013b} and the known single WR populations of the Large Magellanic Cloud (LMC) from \citet{massey2002} using the conversion to absolute magnitudes by \citet{eldridge2013}.
We also show the magnitude range found by \citet{folatelli2016} for the potential progenitor of iPTF13bvn.
Given the overall magnitude range, we do not distinguish between Johnson-Cousins, Bessel, and \textit{HST} Vega magnitudes for similar wavelengths (e.g., \textit{U} and $F336W$).

\begin{table}
\caption{Variability limits.}
\begin{tabular}{c c c c r c }
\multicolumn{1}{c}{Band}
&\multicolumn{3}{c}{Variability $[10^3\text{L}_\odot]$}
&\multicolumn{2}{c}{Slope [$10^3\text{L}_\odot$/yr]}\\
&\multicolumn{1}{c}{RMS}
&\multicolumn{1}{c}{$\langle\sigma^2\rangle^{1/2}$}
&\multicolumn{1}{c}{Sample}
&\multicolumn{1}{c}{Prog.}
&\multicolumn{1}{c}{Sample}\\
\hline
$R$ & $2.7$ & $0.6$ & $1.8\pm0.7$ & $-0.5\pm0.9$ & $0.3\pm0.3$\\ 
$V$ & $2.0$ & $0.6$ & $1.7\pm0.4$ & $0.8\pm0.5$ & $0.2\pm0.2$\\ 
$B$ & $2.6$ & $0.6$ & $2.1\pm0.9$ & $0.2\pm0.7$ & $0.2\pm0.2$\\ 
$U$ & $2.3$ & $0.5$ & $1.1\pm0.6$ & $-0.3\pm0.7$ & $0.1\pm0.1$\\ 

\end{tabular}
\label{tbl:var}
\end{table}

Next, we analyze the subtracted images prior to \thisevent~to constrain the variability of the progenitor. 
We again use the comparison sample grid to place constraints on the level of any systematic noise. 
Figure \ref{fig:4pre} shows the luminosities of the progenitor along with the twelve comparison positions.
The gray regions again show the root-mean-square (RMS) scatter of the comparison sample about their mean. 
The horizontal dashed lines show their overall mean dispersion about zero. 

We first examine the ``stochastic'' variability of the progenitor using the RMS of the pre-SN difference imaging light curve as compared to the variance predicted from the estimated errors. 
These values are reported as ``RMS'' and ``$\langle\sigma^2\rangle^{1/2}$'' in Table \ref{tbl:var}.
We also determine the average RMS of the comparison sample and its standard deviation which is reported as the first ``Sample'' column in Table \ref{tbl:var}.
The $\sim2500\text{L}_\odot$ RMS of the progenitor appears to be larger than the variance of $\sim600\text{L}_\odot$ predicted by the estimated errors.
However, \texttt{ISIS} tends to underestimate errors because it considers only Poisson uncertainties.
A better estimate of the expected noise are the variances of the comparison sample, which are 2-4 times larger than predicted by the formal uncertainties.
The progenitor's random variability is consistent with these values, so we conclude that there is no significant evidence for ``stochastic'' variability.
We adopt an upper limit on the variability of $\lax2500\text{L}_\odot$ in all four bands.

To investigate any long-term trends in luminosity, we perform a linear fit, $L(t)=At+B$, to the pre-SN light curves of the progenitor with the results summarized in Table \ref{tbl:var}.
The slopes are both positive and negative across the bands, and are on the order of $\sim500\text{L}_\odot$/year with comparable formal errors of $\sim700\text{L}_\odot$/year (reported in Table \ref{tbl:var} as ``Prog'').
This already suggests that there is no evidence of a long-term luminosity trend. 
Furthermore, the $\chi^2/$dof of the fits is $\approx20$ because the variance of the light curves is significantly larger than the formal uncertainties. 
If we rescale the errors to make $\chi^2/\text{dof}\equiv 1$, then the uncertainties on the slope roughly double and the evidence against any significant trend is stronger yet.
We also carried out linear fits to the comparison sample and report the average of the absolute values of their slopes and their standard deviations as the second ``Sample'' column in Table \ref{tbl:var}.
The slopes found for the progenitor are consistent with both the comparison sample and zero, leading us to conclude that we did not detect any long-term variability of the progenitor in its final years at the level of $|A|\lax1000\text{L}_\odot/\text{year}$.

\section{Discussion}
\label{sec:discussion}

As shown in Figure \ref{fig:limits}, our limits for \thisevent~are the tightest ever obtained for a Type Ibc SN in the \textit{U} and \textit{V} bands, and are comparable to the strongest existing limits in both \textit{B} and \textit{R}.
This demonstrates the power of ground-based difference imaging for the study of ccSNe and their progenitors. 
The limit we place on the \textit{U} magnitude essentially rules out all of the \citet{groh2013b} single star progenitor models. 
The hottest \citet{yoon2012} models might marginally evade these limits although they report only estimates of $M_v$.
Moreover, the formal limit we adopted for this band is very conservative, as the value is driven by the substantial residuals from nearby stars (see Figure \ref{fig:4panelU}). 
Nearly all WR stars from the LMC in the \citet{massey2002} sample are excluded by this limit as well.

These facts are most easily interpreted as support for a binary origin for \thisevent.
\citet{sukhbold2016} also find that the end of life masses of their wind-stripped progenitors tend to be too high to produce the observed light curves of Type Ibc SNe. 
\citet{dessart2011} could reproduce Ibc light curves only if the ejecta mass was $\sim4\text{M}_\odot$, which also likely requires a dominant binary channel for producing Type Ibc SNe. 

We also find that the progenitor could have had very little variability in the $\sim$4 years prior to its explosion, with strong limits on both the random variability (RMS$\lax2500\text{L}_\odot$) and the long-term variability ($|A|\lax1000\text{L}_\odot/\text{year}$).
Since we did not detect the progenitor, our findings are still consistent with a high fractional variability in the observed bands.
However, the absolute variability scale is tiny compared to the bolometric luminosity of a typical WR star \citep[$\gax10^5\text{L}_\odot$, e.g.,][]{groh2013b}.
This lack of eruptive variability shortly before the SN is consistent with prior LBT results for the Type IIb SN 2011dh \citep{szczygiel2012} and the Type IIP ASASSN-16fq \citep{kochanek2017}.

\section{Acknowledgments}

We thank T. Sukhbold and the referee of this paper for useful comments. 
CSK is supported by NSF grant AST-1515876 and AST-1515927. 
This work is based on observations made with the Large Binocular Telescope. 
The LBT is an international collaboration among institutions in the United States, Italy, and Germany. 
The LBT Corporation partners are: the University of Arizona on behalf of the Arizona university system; the Istituto Nazionale di Astro. 
This research has made use of the NASA/ IPAC Infrared Science Archive, which is operated by the Jet Propulsion Laboratory, California Institute of Technology, under contract with the National Aeronautics and Space Administration.

\vfill\eject

\end{document}